# Benchmarking Performance of Deep Learning Model for Material Segmentation on Two HPC Systems

Warren R. Williams[1], S. Ross Glandon[1], Luke L. Morris[2], Jing-Ru C. Cheng[1]
[1]Information Technology Laboratory, US Army Engineer Research and Development Center, Vicksburg, MS 39180
[2]University of Florida, Department of Computer & Information Science & Engineering, Gainesville, FL 32611

**Abstract**
Performance Benchmarking of HPC systems is an ongoing effort that seeks to provide information that will allow for increased performance and improve the job schedulers that manage these systems. We develop a benchmarking tool that utilizes machine learning models and gathers performance data on GPU-accelerated nodes while they perform material segmentation analysis. The benchmark uses a ML model that has been converted from Caffe to PyTorch using the MMdnn toolkit and the MINC-2500 dataset. Performance data is gathered on two ERDC DSRC systems, Onyx and Vulcanite. The data reveals that while Vulcanite has faster model times in a large number of benchmarks, and it is also more subject to some environmental factors that can cause performances slower than Onyx. In contrast the model times from Onyx are consistent across benchmarks.

## 1. Introduction

The demand for intelligent devices and tools that will facilitate safer work environments, safer roadways, and an increased quality of life is ever growing. High-Performance Computing (HPC) systems are powerful tools that aid in the development of Machine Learning (ML) models and applications that seek to meet the goals laid out by this increasing demand. Therefore benchmarking the performance of these HPC systems in utilizing ML methods is essential. These benchmarking results aid in the selection of systems for different projects, as the growing variety in architecture and methodology in ML allows for different systems to meet different needs. These results can also be used for the optimization of systems to improve their performance and allow their schedulers to make more informed decisions when allotting resources. To further benchmarking efforts in this paper, we compare the results of two HPC systems using a Material Segmentation ML performance benchmark.

Benchmarking ML performance is a broad area of research. For example, Thiyagalingam *et al.* developed scientific ML benchmarks to further benchmarking analysis efforts [1]. Malakar *et al.* used ML methods for modeling performance and benchmarking to further efforts to improve HPC scheduling and performance in scientific applications [2]. AIPerf was developed by researchers to provide more informed benchmarking for ML & AI needs on HPC systems [3]. MLPerf is a benchmarking method that targets ML inferencing systems and establishes best practices for ML benchmarking [4].

Material recognition aids intelligent devices, like autonomous vehicles, in their decisions about navigating and interacting with their environment. Material Segmentation (MS) is a method of material recognition that breaks down images and identifies the material present in each of the individual segments. This is most often accomplished with the use of ML models. MS research has been done to aid in identifying the material make-up of the street-level surroundings of a vehicle [5]. Other MS research has been done on the material recognition of multi-view satellite imagery, using the additional information from the multiple images to improve performance, as well as on improved learning methods to support the remote, autonomous handling of nuclear waste during the disposal process by robots [6, 7].

The remainder of this paper is organized as follows: Section 2 will discuss the HPC systems that were benchmarked and the dataset we used throughout our testing. Section 3 provides details on the ML architecture used and the implementation of our benchmark. Section 4 will report and discuss the benchmarking results. Section 5 will contain our drawn conclusions, ideas for future improvements and work of our benchmark, and ideas for further investigations into the subject HPC systems.

## 2. Materials and Methods

The two HPC systems benchmarked in this paper are hosted at the U.S. Army Engineer Research and Development Center (ERDC) DoD Supercomputing Resource Center (DSRC). The selected systems are Onyx and Vulcanite. Onyx is an XC40/50 Cray system, and its node configurations are shown in Table 1. Vulcanite is a Linux cluster that was developed to support research on the development and deployment of GPU-enabled capabilities. Vulcanite is a SuperMicro system that supports a mixture of GPU node configurations as shown in Table 2. Benchmarks for Onyx were performed on the GPU Accelerated nodes. Benchmarks for Vulcanite were performed on 2-GPU, 4-GPU, and 8-GPU nodes, with emphasis on the 4-GPU nodes.

**Table 1: Onyx Node Configurations.**

| Node Configuration | Login | Standard | Large-Memory | KNL | GPU | 2-MLA | 10-MLA |
|---|---|---|---|---|---|---|---|
| Total Nodes | 12 | 4,810 | 4 | 32 | 32 | 60 | 4 |
| Processor | Intel E5-2699v4 Broadwell | Intel E5-2699v4 Broadwell | Intel E5-2699v4 Broadwell | Intel 7230 Knights Landing | Intel E5-2699v4 Broadwell | Intel 6148 Skylake | Intel 6148 Skylake |
| Processor Speed | 2.8 GHz | 2.8 GHz | 2.8 GHz | 1.3 GHz | 2.8 GHz | 2.4 GHz | 2.4 GHz |
| Sockets / Node | 2 | 2 | 2 | 1 | 1 | 2 | 2 |
| Cores / Node | 44 | 44 | 44 | 64 | 22 | 40 | 40 |
| Total CPU Cores | 264 | 211,640 | 176 | 2,048 | 704 | 2,400 | 160 |
| Useable Memory / Node | 247 GB | 121 GB | 1 TB | 90 GB | 247 GB | 172 GB | 735 GB |
| Accelerators / Node | None | None | None | None | 1 | 2 | 10 |
| Accelerator | n/a | n/a | n/a | n/a | NVIDIA P100 PCIe | NVIDIA V100 SXM2 | NVIDIA V100 PCIe |
| Memory / Accelerator | n/a | n/a | n/a | n/a | 16 GB | 16 GB | 32 GB |
| Storage on Node | None | None | None | None | None | None | None |
| Interconnect | Ethernet | Cray Aries | Cray Aries | Cray Aries | Cray Aries | InfiniBand | InfiniBand |
| OS | SLES | CLE | CLE | CLE | CLE | CentOS | CentOS |

**Table 2: Vulcanite Node Configurations.**

| Node Configuration | | | | |
|---|---|---|---|---|
| | Login | 2-GPU | 4-GPU | 8-GPU |
| Total Nodes | 2 | 26 | 8 | 5 |
| Processor | Intel 6126T Skylake | Intel 6126T Skylake | Intel 6136 Skylake | Intel 8160 Skylake |
| Processor Speed | 2.6 GHz | 2.6 GHz | 3.0 GHz | 2.1 GHz |
| Sockets / Node | 1 | 1 | 2 | 2 |
| Cores / Node | 12 | 12 | 24 | 48 |
| Total CPU Cores | 24 | 312 | 192 | 240 |
| Useable Memory / Node | 206 GB | 206 GB | 384 GB | 764 GB |
| Accelerators / Node | None | 2 | 4 | 8 |
| Accelerator | n/a | NVIDIA V100 PCIe | NVIDIA V100 SXM2 | NVIDIA V100 SXM2 |
| Memory / Accelerator | n/a | 32 GB | 32 GB | 32 GB |
| Storage on Node | None | 2 TB NVMe | 4 TB NVMe | 8 TB NVMe |
| Interconnect | EDF InfiniBand 1x | EDR InfiniBand 1x | EDR InfiniBand 2x | EDR InfiniBand 4x |
| OS | RHEL7 | RHEL7 | RHEL7 | RHEL7 |

The dataset used in the benchmark is the MINC-2500 that has been made publically available by Bell *et al*. The full MINC dataset contains over 3 million material samples patches of 23 material categories, and it was developed with the goal of allowing MS and material recognition in real world scenarios. Real-world environments contain more clutter and variation, which created difficulties for models trained on previous datasets. The MINC-2500 contains 2500 samples for each of the 23 categories within MINC. All of the samples in the MINC-2500 have been sized to 362 x 362 pixels [8].

## 3. Material Segmentation

During their research developing the MINC, Bell *et al.* also trained ML models for MS implementation [8]. These models were implemented using the Caffe framework, which has reached its end-of-life. An updated version of this framework, Caffe2, was released, but it has now been merged with PyTorch [9]. Through our efforts to implement one of these models from Bell *et al.*, we chose to convert the model to the PyTorch framework. This conversion would allow for the utilization of continued support and development from PyTorch. This conversion to PyTorch would also allow for the model to take advantage of improvements in GPU hardware. The MS model using the GoogLeNet architecture was chosen to be converted. The GoogLeNet architecture is shown in Figure 1. The conversion was done using the Microsoft Research's Model Management deep neural network (MMdnn) framework converter toolkit [10].

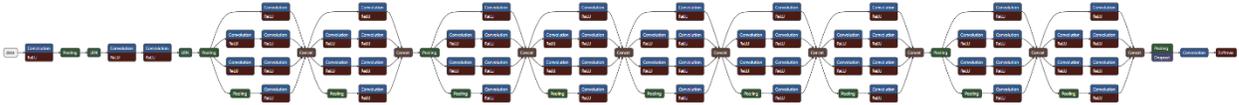

**Figure 1:** *GoogLeNet Neural Network Architecture.* **This diagram was automatically generated by Netron neural network visualizer.**

The benchmark pipeline begins by checking the size of the dataset specified by an argument presented in its execution command. It searches through the full dataset, including subdirectories, to obtain the number of images it has to process. It uses this information to track its progress throughout its execution. During interactive sessions, it displays progress bars showing its progress and performance, from elapsed time to number of iterations per second (each image processed is an iteration). It then works through the dataset one subdirectory at a time. This methodology allowed for the benchmark to be performed on the entire MINC-2500 dataset at once. The images in the MINC-2500 are separated into subdirectories for each of the 23 material categories. This set up also allows for the benchmark to be performed with alternative datasets should that be desired.

The program takes each input image and scales it three times. The image scales used in our testing are one at its original scale, one scaled down by a factor of one over the square root of two (1/√2), and one scaled up by a factor of the square root of two (√2). These are also the scales that the benchmark defaults to use, but there is an argument to specify alternative scales. For each of these scaled images, the program creates a series of smaller windows of the image by cropping them. These collected windows are then passed over to the GPU and handed to the neural network. The model processes them and returns 23 probability maps for each image.

The original MS program then scaled all of the returned probability maps to equivalent dimensions, and the maps were then combined to give 23 cumulative probability maps. These cumulative maps were then used to perform Dense Conditional Random Field inference. This inference was performed a set number of times to yield the output segmentation. However, as our initial focus for the benchmark was on GPU performance, this section of the program is not considered in the current benchmark. Re-implementing this may be a future direction for the benchmark. Alternatively another method of inferencing could be implemented in the future instead. The method proposed by Baqué *et al.*, could offer increases to inferencing speeds [11]. Other alterations to the benchmark could include attempts to perform more of the pre or post processing on the GPU.

The benchmark measures and records the execution time of the model's neural network for each image. It also records the total time it takes to process each directory. The average model execution time is calculated for each directory as well. When the average model time for the directory is calculated, we exclude the first execution time. We observed the first measured time is always abnormally long, as it includes the setup time that occurs the first time that the model is called.

The total time to process each directory and the average model time for each directory are printed to the screen and recorded into a text file. All of the individual model times are recorded into a CSV file as well, where they are separated into rows by directory. Once the program has gone through all of the provided images and directories, it will then calculate an average model time for the entire execution. This total average model time includes all of the individual model times except for the first time in each directory. This total average model time is also stored in the text file.

## 4. Results

Benchmark data was gathered across 5 runs on Onyx GPU nodes. The benchmark data gathered for Vulcanite spans 6 runs on 4 GPU nodes, 3 runs on 2 GPU nodes, and 1 run on an 8 GPU node. The benchmark was performed with some variation to its configurations. The cumulative results on model execution times from all runs are shown in Figure 2. The Nvidia V100 GPUs on the Vulcanite nodes are the next step up in performance on Nvidia's Tesla GPU line from the Nvidia P100 GPUs that are on the Onyx nodes. The mean, average median, and average $5^{th}$ percentile model times on both systems reflect that performance as expected. However the model times on each system begin to show an inverse result as you get closer to their max times. The average $95^{th}$ percentile model times for Vulcanite nodes start to become slower than the model times on Onyx nodes of the same percentile.

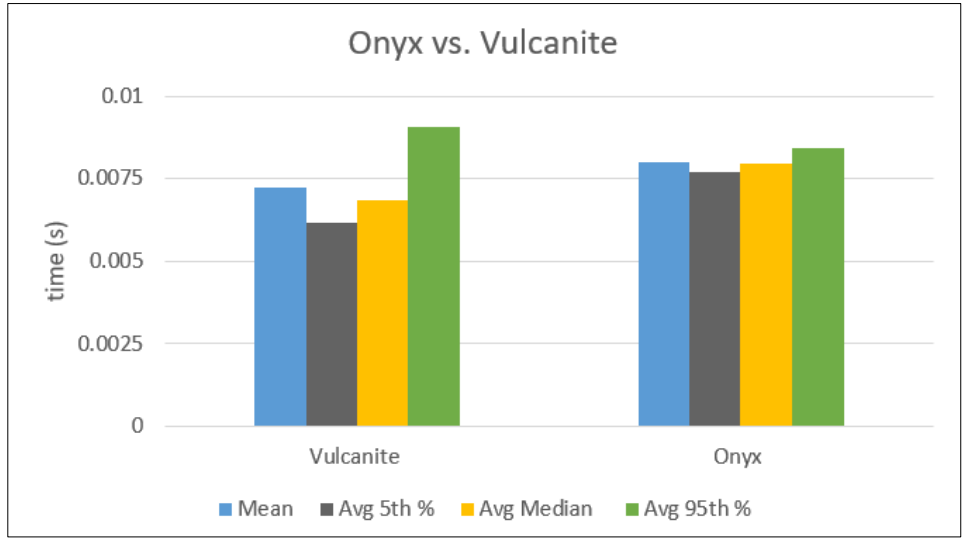
**Figure 2:** *Onyx vs. Vulcanite.* **System model times from multiple benchmarks are represented.**

This unexpected result prompted further investigation into our data. Figure 3 shows the average model times for a selection of the benchmarking runs on the two systems. The performance of the Onyx nodes appear to be more consistent across benchmarking runs at different times than the performance of the Vulcanite nodes. However the majority of the Vulcanite benchmarks do perform with average model times faster than the Onyx average model times. This fits with the expectation of its higher performing GPUs, but some runs appear to be affected by periods of low performance that cause the mean time to rise above Onyx's consistent performance.

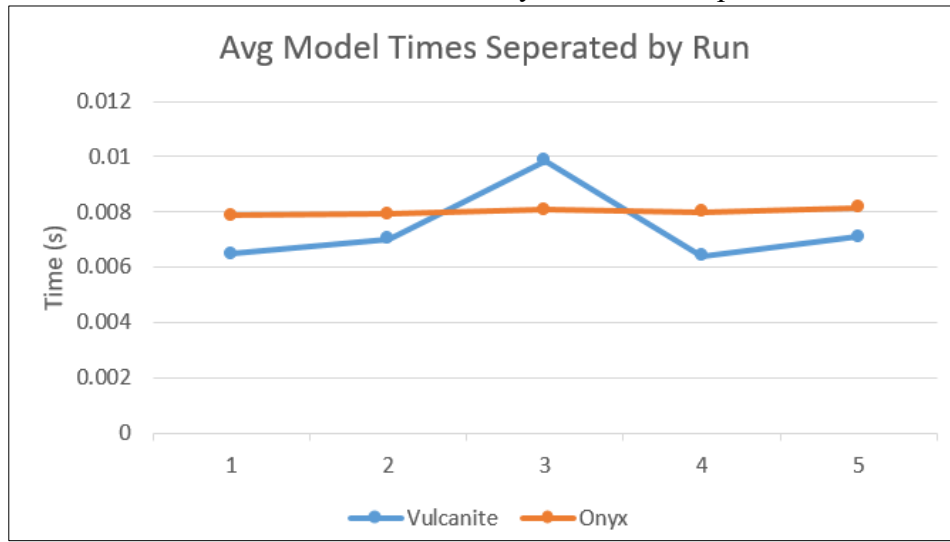
**Figure 3:** *Average Model Times Separated by Run.* **Select benchmark runs are shown here that represent the results of all benchmarks gathered.**

To acquire more insight into the difference between these runs, individual model times for benchmarking runs were graphed temporally. Figure 4 shows a selected subsets of these model times. The graph on the right is from a benchmark run where the Vulcanite node displayed the expected performance and the graph on the left is from a benchmark run that demonstrated the lower performance. These graphs provide important insight into the variation of behavior between these low performance runs and the runs that perform as expected on Vulcanite. On both graphs you can see what would appear to be two baseline performances. The lower performance stays mainly on the slower line, accelerating to the faster times only occasionally, and the normal performance inversely stays at this higher speed most of the run, only slowing to the secondary line for brief periods. It is reminiscent of a change to the GPU's frequency, as though it is throttling down to help cool the GPU down from higher temperatures.

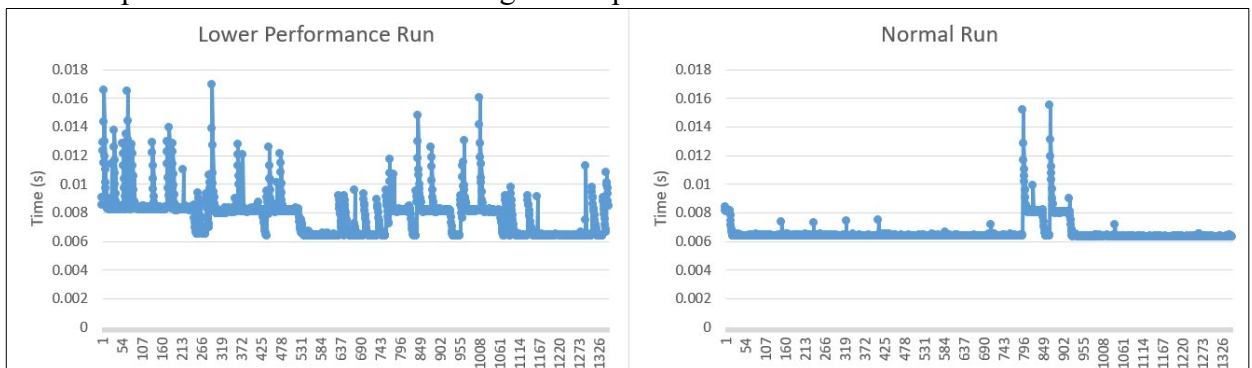

**Figure 4:** *Vulcanite Low Performance Run vs. Normal Run.* **These are individual model inference times graphed on a line in the order they were gathered.**

It is relevant to note that while Vulcanite's performance displayed some periods of lower performance, these times are still incredibly fast. In most applications the difference between 0.007s and 0.009s is negligible. This is unlikely to be the factor that is slowing down the performance of many ML applications. Also it is not the case most of the time. In most of the benchmarks, Vulcanite was processing at the higher speeds and outperforming the Onyx benchmarks. Vulcanite's benchmarks outperformed Onyx's as a whole until we get to model times in the 90th percentile. However Onyx does provide more consistency in its performance, which is an important factor to consider in some applications.

The heat generated from the node due to running the benchmark would likely be similar between the two nodes. There is a possibility of issues within specific nodes, but this lower performance was experienced on multiple nodes, so that likely is not the case. That then leaves the possibility that it is being affected by the activity on the neighboring nodes. If the surrounding nodes are running high demand jobs and the heat is affecting the benchmarking nodes performance, that could explain why the performance is so notably different.

Further investigation into the area in which Vulcanite is placed within the data center environment led to discovering that Vulcanite is entirely air cooled. This leaves it more susceptible to temperature variations affecting performance. The air surrounding the cold side of Vulcanite also appeared to be a mix of hot and cool air that would lower its ability to efficiently cool down its computing nodes. All of these factors seem to indicate that environmental temperatures are affecting the Vulcanite's performance. Onyx is a liquid cooled system, and liquid cooling is less susceptible to environmental factors. This could explain the more consistent results received from Onyx.

Investigation into the factors affecting this temperature could be done. Such as experiments in which we have control of neighboring nodes and can control varying the load on those to see what affect that has on a node's benchmark. As well as investigation into whether the placement vertically in the rack increases the likelihood of slower performance due to heat traveling up through the system. Also evaluating the thermal performance of the data center space that houses the systems using methods like those reviewed by Gong *et al.* could aid in identifying the main cause of the warmer air temperature around Vulcanite [12].

## 5. Conclusions

Material Segmentation is a material recognition method that breaks images down into smaller segments for processing, often utilizing Machine Learning models. The developed benchmark utilizes a material segmentation machine learning model to analyze the MINC-2500 dataset. The benchmark gathers the model execution times as the dataset is processed. The resulting performance data for Onyx and Vulcanite displayed unexpected variation in Vulcanite's benchmark times. Through further investigations of the results and the system environments, it appears that Vulcanite may have performances that are negatively impacted by increased temperatures around and within the system.

Future development of the benchmark program could be done to re-implement the currently unconsidered pieces of the MS process to perform the Dense Conditional Random Field inference and return output probability maps. Dense CRF is notoriously slow, so adding it to the benchmark would have a significant effect on performance. Implementing a GPU profiler to monitor the GPU state while benchmarking would also be interesting and helpful in identifying if temperature throttling is in fact the cause of the decreased performance. Nvidia's nvml tool could be used to produce a profile and record the state of the GPU throughout the benchmark [13]. Utilizing an alternative dataset like RELLIS-3D within the benchmark would also be interesting if there was any effect on the models performance and output [14].

**Acknowledgements**

The DoD High Performance Computing Modernization Program (HPCMP) and the HPC Internship Program provided the fundamental support and hosting of resources for this project